\documentclass[12pt,preprint]{aastex}
\usepackage{amsmath,emulateapj5,apjfonts,epsfig}
\newcommand{\chandra}{{\it{Chandra}}}
\newcommand{\beppo}{{\it{BeppoSAX}}}

\newcommand{\xmmn}{{\it XMM-Newton}}
\newcommand{\rvir}{r_{\rm{vir}}} 

\newcommand{\msun}{{\rm M}_{\sun}}
\newcommand{\lsun}{{\rm L}_{\sun}}

\newcommand{\hkpc}{$h_{70}^{-1}$~kpc}
\newcommand{\hmpc}{$h_{70}^{-1}$~Mpc}

\newcommand{\mytable}{tablehere} 
\newcommand{\myfigure}{figure*} 


\submitted{To Appear in the Astrophysical Journal}
\submitted{Received 2002 September 10; Accepted 2002 November 21}
\journalinfo{To Appear in The Astrophysical Journal, Vol. 586, No. 1}

\shorttitle{A2029 Dark Matter Profile}
\shortauthors{Lewis et al.}

\begin{document}

\title{\chandra{} Observations of Abell 2029: The Dark Matter
Profile Down to $<0.01r_{\rm{vir}}$ in an Unusually Relaxed Cluster}

\author{Aaron D. Lewis\altaffilmark{1},  
David A. Buote\altaffilmark{1}, John T. Stocke\altaffilmark{2}}
\email{lewisa@uci.edu, buote@uci.edu, stocke@casa.colorado.edu}
\altaffiltext{1}{University of California, Irvine, Department of
Physics and Astronomy, 4129 Frederick Reines Hall, Irvine, CA,
92697-4575}
\altaffiltext{2}{Center for Astrophysics and Space Astronomy,
University of Colorado, 389 UCB,  Boulder, CO 80309}

\received{2002 September 11}
\accepted{}
\journalid{}{}
\articleid{}{}

\begin{abstract}  
We have used a high spatial resolution \chandra{} observation to
examine the core mass distribution of the unusually regular cD
cluster Abell 2029. This bright, nearby system is especially
well-suited for analysis of its mass distribution under the
assumption of hydrostatic equilibrium: it  exhibits an undisturbed,
symmetric X-ray morphology, and a single-phase intracluster medium
(ICM).  From the deprojected temperature and density profiles we
estimate the total mass, and the contributions of the gas and dark
matter (DM) components from $<3\arcsec$ to $\sim3\arcmin$
($<4.4-260$\hkpc, $0.001-0.1\rvir$). The gas density profile
is not adequately described by a single $\beta$-model fit, due to
an increase in the density at the center ($r<17$\hkpc,
$<12\arcsec$), but it is well fitted by either a double
$\beta$-model, or a ``cusped'' $\beta$-model. The temperature data
increase as a function of radius and are well-fitted by a
\citeauthor{ber86} profile and approximately by a power-law
$T(r)\propto r^{\alpha_T}$, with $\alpha_T=0.27\pm0.01$.

Using the fitted profiles to obtain smooth functions of density and
temperature, we employed the equation of hydrostatic equilibrium to
compute the total enclosed mass as a function of radius. We report
a total mass of $9.15\pm0.25 \times 10^{13} h_{70}^{-1}~\msun$
within $260$\hkpc, using the chosen parameterization of gas density
and temperature. The mass profile is remarkably well described down
to $0.002\rvir$ by the \citeauthor*{nav97} (NFW) profile, or a
\citeauthor{her90} profile, over 2 decades of radius and 3 decades
of mass. For the NFW model, we measure a scale radius $r_s =
540\pm90$\hkpc{} ($\approx0.2\rvir$) and concentration parameter
$c=4.4\pm 0.9$. The mass profile is also well-approximated by a
simple power-law fit ($M(<r)\propto r^{\alpha_m}$), with
$\alpha_m=1.81\pm0.04$ (corresponding to a logarithmic density
profile slope of $-1.19\pm 0.04$). The density profile is too
shallow to be fitted with the profile described by
\citeauthor{moo99}. The consistency with NFW down to $<0.01\rvir$
is incompatible with the flattened core DM profiles predicted for
self-interacting DM (e.g., \citeauthor{spe00}), and thus contrasts
with the strong deviations from CDM predictions observed in the
rotation curves of low surface brightness galaxies and dwarf
galaxies. This suggests that while CDM simulations may adequately
describe objects of cluster mass, they do not currently account
properly for the formation and evolution of smaller halos.

Assuming that the cD dominates the optical cluster light within its
effective radius ($R_e=52\arcsec$, 76\hkpc), we observe a total
mass-to-light ratio $M/L_V \approx12 \msun/\lsun$ at $r<20$\hkpc,
rising rapidly to $>100 \msun/\lsun$ beyond 200\hkpc. The
consistency with a single NFW mass component, and the large $M/L$
suggest the cluster is DM-dominated down to very small radii
($\lesssim 0.005\rvir$). We observe the ICM gas mass to rise from $3
\pm 1\%$ of the total mass in the center to $13.9 \pm 0.4\%$ at the
limit of our observations. This provides an upper limit to the
current matter density of the Universe, $\Omega_{m}\leq
0.29\pm0.03~h_{70}^{-1/2}$.
\end{abstract}


\section{Introduction\label{sec_intro}}

The large mass-to-light ratios ($M/L$) of galaxy clusters
\citep[$M/L_B \approx 200-300 h_{70}~\msun/\lsun$, e.g.,][]{gir02}
indicate that they contain large quantities of dark matter (DM
hereafter). The nature and distribution of DM is the current
subject of much theoretical work in cosmology, with detailed
simulations yielding different expectations for the amount and
distribution of DM in cluster cores
\citep[e.g.,][]{nav97,moo99,dave01}. X-ray observations of the
density and temperature of the hot intracluster medium (ICM) gas
probe the mass of a galaxy cluster, under the assumption of
hydrostatic equilibrium . Such data potentially provide constraints
on cluster DM simulations, and thus test DM theory
\citep[e.g.,][]{evr96,ara02,san02}.

Prior generations of X-ray satellites have provided a wealth of
cluster observations, from which we have begun to build a picture
of large-scale radial temperature variations
\citep[e.g.,][]{irw01,ett02b}, as well as two dimensional
temperature maps of disturbed systems
\citep[e.g.,][]{bri94,mar98,mar99,deg99b,joh02}. However, detailed
temperature and density profiles at the smallest scales ($r<0.1
\rvir$) exist for only a few systems, such as Virgo \citep{nul95},
which exhibit irregularities in their cores. While gravitational
lensing studies provide a unique and important probe of DM in
cluster cores \citep{dah02,nat02,san02}, they generally cannot be
applied to nearby systems, and they may also be contaminated by
other sources of mass along the line of sight. The advent of the
\chandra{} and \xmmn{} satellites now allows us to measure the ICM
properties with simultaneous spatial resolution comparable to
optical studies, thus providing a completely independent mass
estimator over the same spatial scales. The main criticism levelled
at X-ray studies is the veracity of the hydrostatic equilibrium
assumption in the actual clusters under study. Several groups have
now obtained X-ray constraints on the DM profiles of clusters of
galaxies, which are apparently consistent with the CDM paradigm
\citep[e.g.,][]{dav01,pra02,ett02a,sch01,all01,mat02,ara02}. Such
systems are either too distant for a detailed analysis at
$<0.1\rvir$, or they contain morphological disturbances indicating
possible departures from hydrostatic equilibrium, especially at
$r\lesssim 100$\hkpc. Although simulations suggest that the X-ray
analysis of such clusters is generally unaffected
\citep{tsa94,evr96,mat99}, some authors argue that mass estimates
will be biased at small radii in such systems, which may reflect
the majority of clusters \citep[see especially][]{mar02}.

Abell 2029 (A2029) is a nearby ($z=0.0767$), well-studied cluster
of galaxies which provides an excellent opportunity to probe the
dark matter content of a massive object. It has a very high X-ray
flux and luminosity, as well as a hot ICM, indicating a massive
system. We have previously analyzed the \chandra{} observations of
A2029 in \citet*[][Paper 1 hereafter]{P5}, presenting the
temperature and Fe abundance data. It exhibits a very regular
optical and X-ray morphology, and is an excellent example of a
relaxed system with no evidence of disturbances (e.g., shock
fronts, filaments, or ``cold fronts'') present in other systems. In
a morphological analysis of 59 clusters, \citet{buo96b} found the
global X-ray morphology of A2029 to be among the most regular in
the sample. Though it contains a wide angle tail (WAT) radio
source, there are no coincident X-ray ``holes'', such as those
found in the clusters Hydra A \citep{nul02}, and Perseus
\citep{fab02}. There is no optical or X-ray spectroscopic evidence
for a cooling flow, though the X-ray temperature drops to $2-3$keV
in the central $5\arcsec$ (7\hkpc). Thus, this system is almost
uniquely well-suited for analysis of its mass distribution since
the hydrostatic equilibrium assumption should apply with high
accuracy.

In the current paper we present estimates of the gas density and
temperature profiles (\S \ref{sec_obs}), the total mass and the DM
density profile (\S \ref{sec_mtot}), as well as the gas mass (\S
\ref{sec_gas}). In \S \ref{sec_dis} we make a comparison with the
stellar mass profile. We discuss the implications of our analysis
and present our conclusions in \S \ref{sec_conc}. Throughout this
paper, we assume a cosmology of
H$_0=70$~$h_{70}$~km~s$^{-1}$~Mpc$^{-1}$, $\Omega_{m}=0.3$, and
$\Lambda=0.7$, implying a luminosity distance to A2029 of
347~\hmpc{} and an angular scale of 1.45 kpc~arcsec$^{-1}$.

\section{Observations and Data Reduction\label{sec_obs}}

A2029 was observed by the \chandra{} observatory for 20ks on the
ACIS S3 chip at a focal plane temperature of -120 C (see Paper 1
for details). We have reanalyzed the data in a similar manner to
Paper 1, but using the latest version of the \chandra{} calibration
(CALDB 2.15). Briefly, the data were processed to mitigate the
effects of charge transfer inefficiency (CTI) using the
ACISCtiCorrector.1.37 software\footnote{Available from the
\chandra{} contributed software page at
http://asc.harvard.edu/cont-soft/soft-exchange.html}
\citep{tow00}. We then subtracted the available source-free
extragalactic sky background maps\footnote{The background maps have
been CTI-corrected in exactly the same manner as the source data.}
using the make\_acisbg software created by Maxim Markevitch
\citep[see, e.g.][]{mar00}. We have also applied the latest
correction to the ARF files to account for an apparent
time-dependent degradation in QE (using the ``corrarf''
routine\footnote{See
http://cxc.harvard.edu/cal/Links/Acis/acis/Cal\_prods/qeDeg/
index.html\label{foot_cor} for a description of the effect and
links to the software.}). This correction has a small but
significant effect on our fitted temperatures at larger radii,
resulting in systematically lower values than our previous analysis
(see \S\ref{subsec_sys} regarding the effect on the mass profile).

\subsection{Binning and Spectral Analysis\label{subsec_bin}}

We extracted spectra in concentric, circular annuli centered on the
peak of the X-ray emission. We have used a different set of annuli
than that used in Paper 1, to optimize the constraints on the shape
of the mass profile in the center as well as the slope of the
temperature profile at all radii. This results in a total of 7
annuli, though this choice does not affect our results (see \S
\ref{subsec_sys}).

Our primary analysis relies on spectra in the energy range 0.7-8.0
keV, but we have explored the effects of using a lower limit of 0.5
or 1.0 keV (see \S \ref{subsec_sys}). Using {\sc xspec}, we fit the
extracted spectra with the {\sc apec} plasma emission model
absorbed by Galactic neutral hydrogen. We adopt the
weighted-average Galactic value of N$_H = 3.14 \times
10^{20}$~cm$^{-2}$ obtained from the W3N$_H$ HEASARC tool. The
model normalization (from which we calculate the electron density,
$n_e$, and gas mass density, $\rho_g$), the gas temperature,
$T_{g}$, and the Fe abundance were allowed to be free parameters in
the fits, with all other elements tied to Fe in their solar ratios.
To properly recover the three-dimensional properties of the X-ray
emitting ICM, we have performed a spectral deprojection analysis
using the {\sc xdeproj} code of \citet{buo00c}. We start at the
outside working our way to the core in an ``onion-peeling'' method
which accounts for the cumulative projected emission from the outer
annuli. For details of our deprojection technique see
\citet{buo00c} and \citet{P7}. To estimate the uncertainties on the
fitted parameters we simulated spectra for each annulus using the
best-fitting models and fit the simulated spectra in exactly the
same manner as the actual data. From 100 Monte Carlo simulations we
compute the standard deviation for each free parameter which we
quote as the ``1$\sigma$'' error.

\subsection{Azimuthally Averaged Gas Density and
Temperature\label{subsec_rhot}}

\begin{\myfigure}[ht]
\parbox{0.49\textwidth}{
\centerline{\epsfig{file=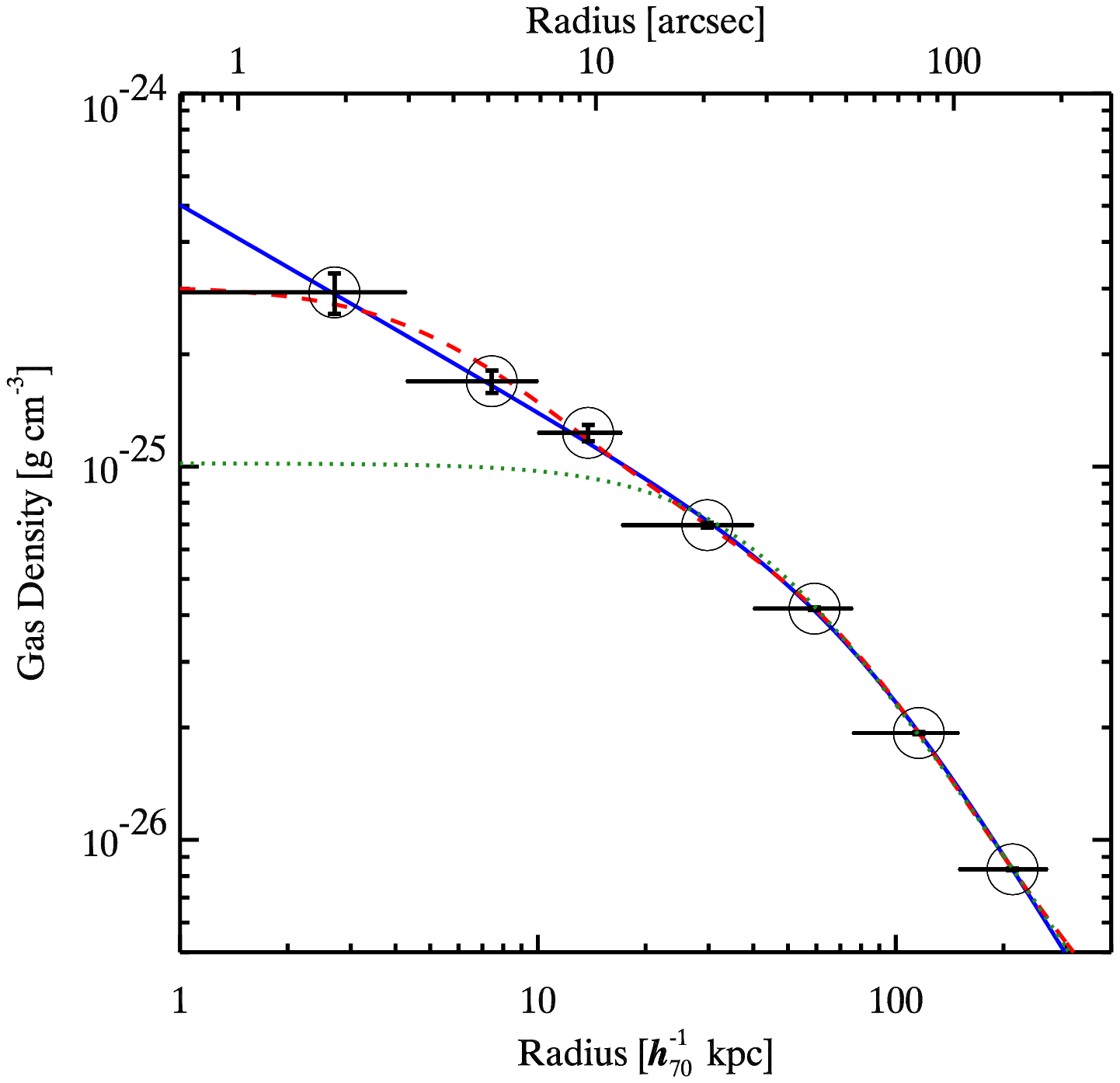, scale=0.59}}
}
\parbox{0.49\textwidth}{
\centerline{\epsfig{file=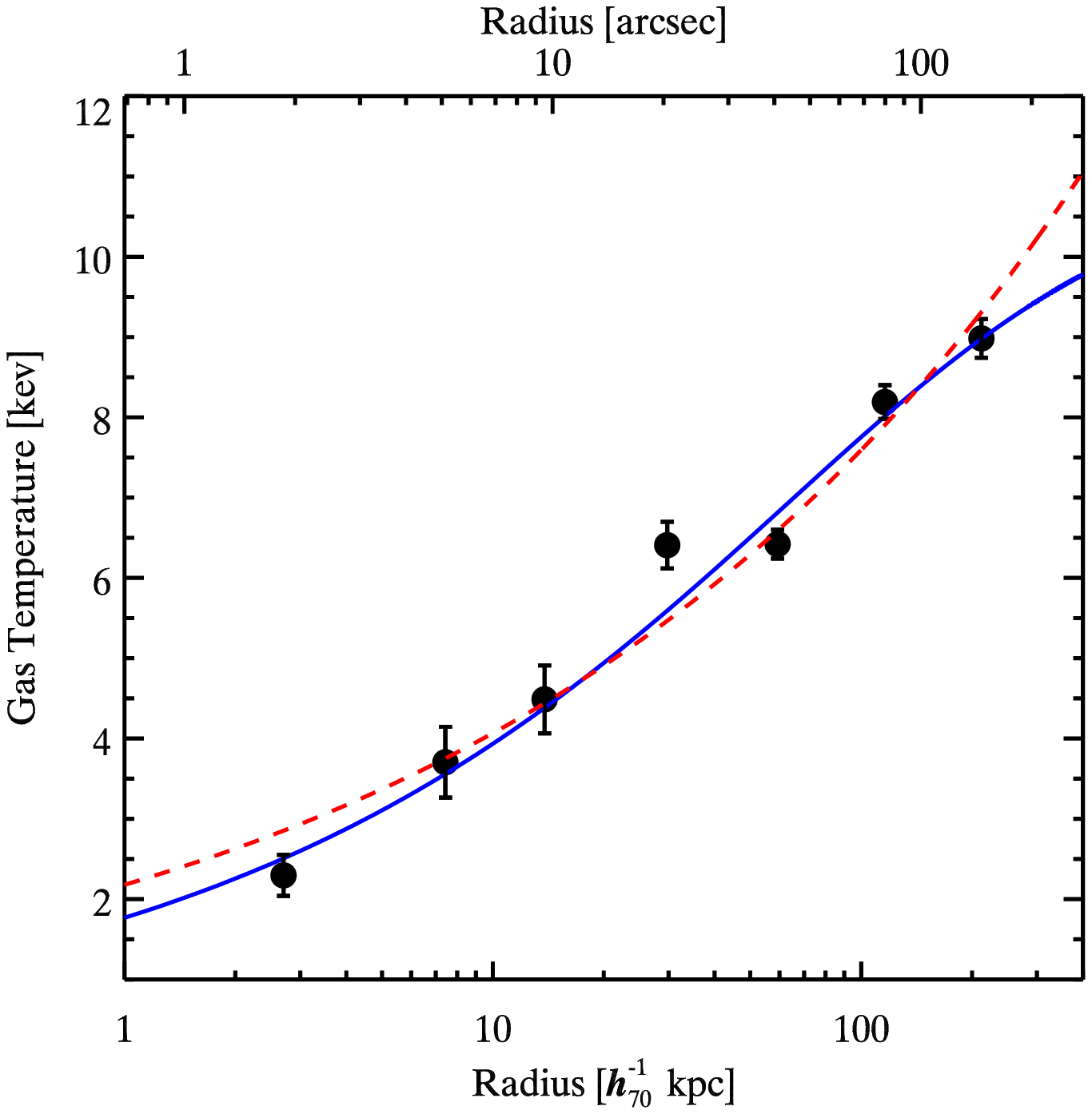, scale=0.59}}
}
\caption{\emph{Left Panel}: \chandra{} radial gas density profile
of A2029. For clarity, large open circles are centered on the data
points (the smallest error bars are difficult to see in the
logarithmic scaling). Horizontal bars indicate the sizes of the
annuli used to extract spectra, and the limits of the spherical
shells in our de-projected analysis. Overlaid are the best-fitting
cusp model (\emph{solid curve}), single $\beta$-model (\emph{dotted
curve}), and double $\beta$-model (\emph{dashed curve}).
\emph{Right Panel}: \chandra{} radial temperature profile of A2029.
Overlaid are the best-fitting \citeauthor{ber86} model (\emph{solid
curve}), and simple power-law model (\emph{dashed curve}).
}
\label{fig_rho_t}
\end{\myfigure}

In Figure \ref{fig_rho_t} we present the radial gas density and
temperature data for A2029 (left and right panels, respectively).
The horizontal bars indicate the widths of the annuli used, and are
not error bars. Note the very small statistical errors in both
$\rho_g$ and $T_g$, which indicate the precise spectral constraints
obtained from these data. Since we wish to fit smooth,
parameterized functions to the data (see below), we choose to
assign an emission-weighted effective radius, $\bar{r}$, to each
annulus $i$ \citep[see, e.g.,][]{mcl99}:
\begin{equation}
\bar{r}_i = [(r_{out_i}^{3/2} +  r_{in_i}^{3/2})/2]^{2/3}.
\label{eq_re}
\end{equation}

While $\bar{r}$ should strictly be iteratively calculated from the
fitted density profile, it is easy to show that eq. \ref{eq_re} is
an excellent approximation to the correct $\bar{r}$ over the
entire profile \citep[e.g.,][see also \S \ref{subsec_sys}]{ett02b}.

\section{Total Enclosed Mass Estimate\label{sec_mtot}}

We make the assumptions of hydrostatic equilibrium and spherical
symmetry, such that the enclosed mass is
\begin{equation}
M_{tot}(<\bar{r}) = -{\frac{\bar{r}kT_g}{G\mu
m_p}}\biggl[\frac{d{\rm ln}\rho_g}{d{\rm ln}r} +
\frac{d{\rm ln}T_g}{d{\rm ln}r}\biggr],
\label{eq_mtot}
\end{equation}
\noindent
where $k$ is Boltzmann's constant, $G$ is the constant of
gravitation, $\mu$ is the mean atomic weight of the gas (taken to
be 0.62), and $m_p$ is the atomic mass unit. To obtain the
instantaneous logarithmic derivatives necessary to evaluate eq.
\ref{eq_mtot}, we fit parameterized models to both the density and
temperature. By parameterizing the $\rho_g$ and $T_g$ data we
derive a mass profile that may be smoother than the true
mass distribution. This approach, therefore, is best suited for
interpreting average properties of $M(<r)$, such as its radial
slope and comparison with DM simulations (also smooth), which are
the focus of the present paper. Key advantages of this method are
that it is simple to implement, and the mass profile is
straightforward to interpret in terms of the input $\rho_g$ and
$T_g$ profiles.

\subsection{Temperature and Density Profiles\label{subsec_profiles}}

We initially fit the gas density data with the ubiquitous 
$\beta$-model:
\begin{equation}
\rho_{g}(r) = \rho_{g_0}[1+ (r/r_c)^2]^{-3\beta/2},
\label{eq_beta}
\end{equation}
\noindent
where $\rho_{g_0}$ is the central gas density, $r_c$ is the core
radius, and $-3\beta$ is the slope of the profile at $r\gg r_c$. The
result is overlaid on the data as a dotted curve (Fig.
\ref{fig_rho_t}, left panel). Due to a peak in the profile at
$<17$\hkpc{} (the first 3 data points), the $\beta$-model does not
provide an acceptable fit (see Table \ref{tab_rho_t} below).

\begin{\mytable}
\begin{center}
\small
Table \ref{tab_rho_t}: Gas Density and Temperature Fits
\tablecaption{Gas Density and Temperature Fits}
\begin{tabular}{lrclc}
\tableline
\tableline
$\rho_g$-Model	&$(\chi^2$/dof)	&$\beta$	&$r_c$[$\arcsec$]
&$\alpha_{\rho}$ \\
\tableline
cusp	  	& 6.6/3		&$0.54\pm 0.01$	&$53.4\pm  4.4$
&$0.55\pm0.03$ \\
1-$\beta$	& 101.8/4	&$0.48\pm 0.01$	&$26.4\pm  1.1$	&\nodata \\
2-$\beta$	& 2.0/1		&$0.34\pm 0.37$	&$ 3.7\pm 4.0$	&\nodata \\
			&			&$0.76\pm 0.14$	&$53.2\pm 9.5$	&\nodata \\
\tableline
$T_g$-Model	&$(\chi^2$/dof)	&$T_{\infty}$[keV]	&$r_c$[$\arcsec$]
&$\alpha_T$ \\
\tableline
B\&M		& 14.3/4	&$11.1\pm 1.6$	&$122.1\pm 125.3$
&$0.36\pm0.05$ \\
Power		& 19.8/5	&\nodata		&\nodata
&$0.27\pm0.01$ \\
\end{tabular}
\end{center}
\small
{\sc Note.}-- For the cusp model, we find 
$\rho_{g_c} = 6.6\pm 0.8 \times10^{-26}$g cm$^{-3}$. 
For the double-$\beta$ model, $\rho_{g1_0}=3.0\pm
0.5\times10^{-25}$g cm$^{-3}$ and $\rho_{g2_0}=5.6\pm
1.8\times10^{-26}$g cm$^{-3}$.
\label{tab_rho_t}
\normalsize
\end{\mytable}
\smallskip

We explored two additional models: (1) the `cusp' model, which is
a modified $\beta$ model given by
\begin{equation}
\rho_{g}(r) = \rho_{g_c} 2^{3\beta/2 - \alpha_{\rho}/2}
(r/r_c)^{-\alpha_{\rho}}[1+ (r/r_c)^2]^{-3\beta/2 +
\alpha_{\rho}/2},
\label{eq_cusp}
\end{equation}
\noindent
where $\rho_{g_c}\equiv\rho_{g}(r_c)$, and the $\alpha_{\rho}$
parameter allows a steepening of the profile at $r<r_c$, and (2) a
double-$\beta$ model \citep[e.g.,][]{xu98,moh99} given by
\begin{equation}
\rho_{g}(r) = \sqrt{\rho_{g1}^2 + \rho_{g2}^2 },
\label{eq_2beta}
\end{equation}
where $\rho_{g1}$ and $\rho_{g2}$ are each given by eq.
\ref{eq_beta}.

The double-$\beta$ and cusp models both provide satisfactory fits
to the data (dashed and solid curves, respectively, Figure
\ref{fig_rho_t}, left panel), though the reduced $\chi^2$ is
slightly improved for the double-$\beta$ model. We present the
results of the gas density fits in Table \ref{tab_rho_t}. It is
apparent that both the cusp and the double-$\beta$ models are
sensitive to a break at $\approx53\arcsec$, and that the parameters
for the first component of the double-$\beta$ model are not well
constrained. We have chosen the cusp model as our ``reference'' fit
for the rest of our analysis for two reasons: (1) it provides a
similar quality fit with two fewer free parameters than the
double-$\beta$ model; (2) it yields a positive mass in the inner
shell, which the $\beta$ models do not (see \S \ref{subsec_sys}).

\begin{\myfigure}[ht]
\parbox{0.49\textwidth}{
\centerline{\epsfig{file=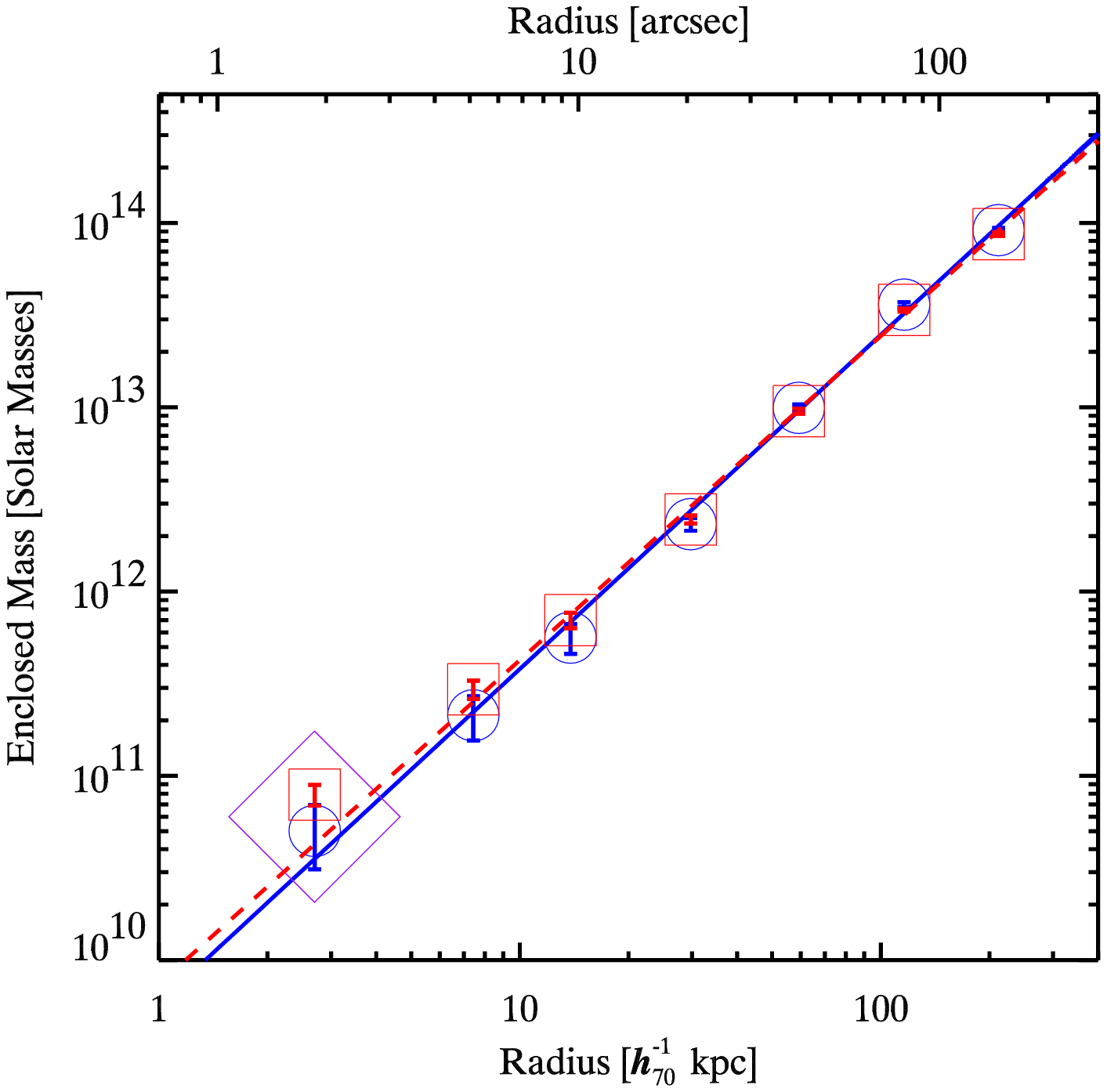, scale=0.57}}
}
\parbox{0.49\textwidth}{
\centerline{\epsfig{file=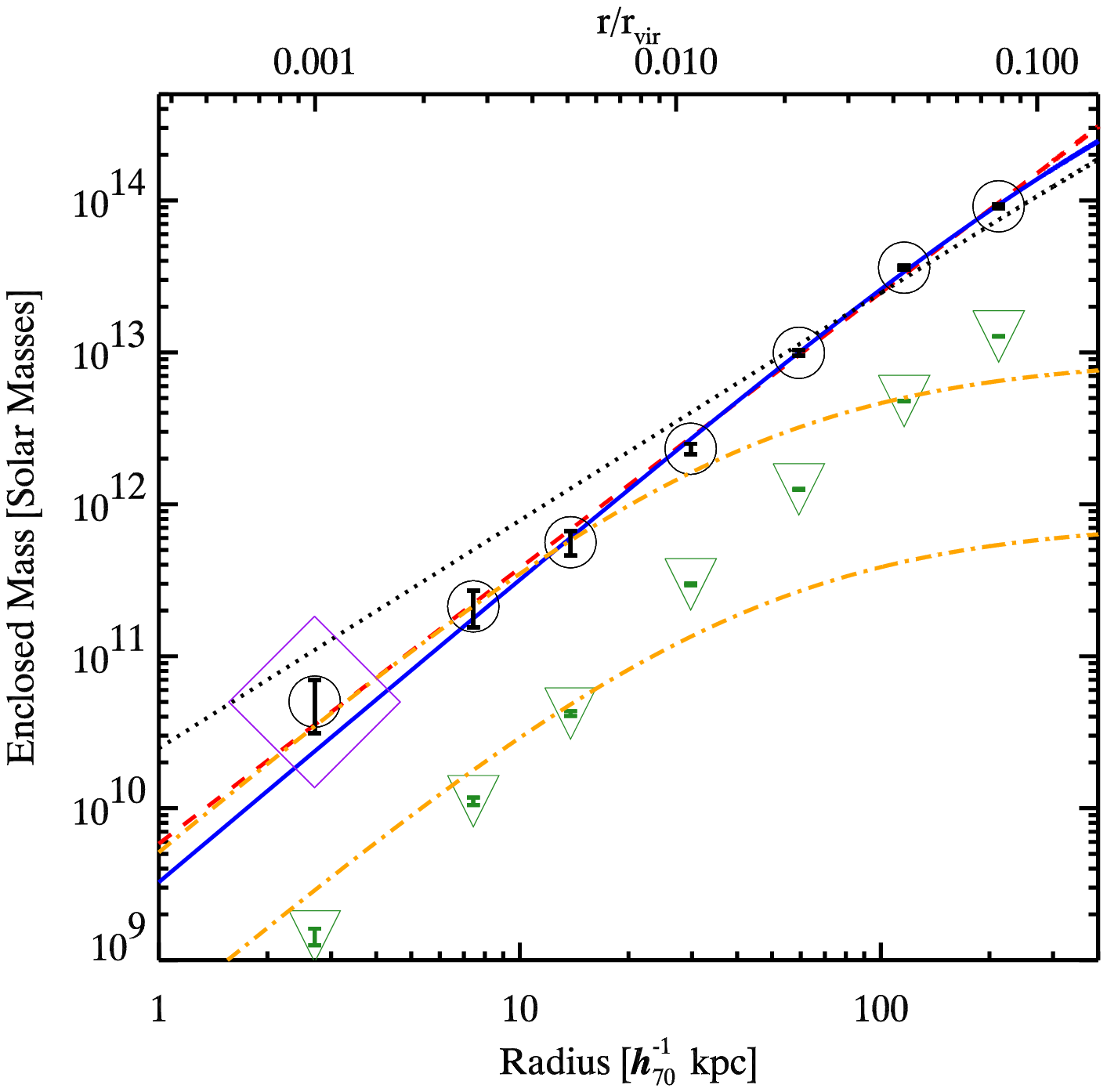, scale=0.57}}
}
\caption{\emph{Left Panel}: Total enclosed cluster mass, obtained
from the B\&M fit (\emph{open circles}) and the power-law fit
(\emph{open squares}) to the temperature data. The cusp model for
$\rho_g$ was used in both cases. Power-law fits to the mass points
are overlaid on both data sets (\emph{solid line}: B\&M $T_g$
model, \emph{dashed line}: power-law $T_g$ model). 
We have used large open symbols to
identify the data points, as some of the error bars are barely
visible in this logarithmic plot. The first data points are also
enclosed with a large open diamond to emphasize the large
additional systematic uncertainty at this radius (see \S\S
\ref{subsec_mass}, \ref{subsec_sys}). \emph{Right Panel}: Total
enclosed cluster mass (data points enclosed with open circles),
overlaid with three different mass models: NFW (\emph{solid
curve}), power-law (\emph{dashed line}), and M99 (\emph{dotted
curve}). 
The total enclosed gas mass is plotted as data points
enclosed with open triangles. We have also overlaid an estimate of
the stellar mass (\emph{dot-dashed curves}, see \S \ref{subsec_stellar}).
The lower curve assumes a $M_*/L_V$
of 1, the upper curve 12. 
}
\label{fig_mass_compare}
\end{\myfigure}

The temperature data are generally increasing, and are fairly
well-fitted by a simple power-law $T(r) = Ar^{\alpha_T}$, with
$\alpha_T = 0.27\pm0.01$ (overlaid as a dashed line). The power-law
fit misses the central data point entirely, and predicts a steeply
rising slope at the outer point. A somewhat better fit to the data
is obtained with a \citet[][B\&M hereafter]{ber86} profile (which
adds an additional free parameter):
\begin{equation}
T(r) = T_{\infty}[r/(r+r_c)]^{\alpha_T},
\label{eq_tber}
\end{equation}
\noindent
where $T_{\infty}$ is the asymptotic temperature at large radii. We
have overlaid the two fitted models to the temperature data in
Figure \ref{fig_rho_t} (\emph{right panel}). The B\&M model
(\emph{solid curve}) provides an improvement in the fit to the data
vs. the power-law model (\emph{dashed line}) at the smallest and
largest radii. The parameters of the fits are given in Table
\ref{tab_rho_t}. From the table we see that the core radius is not
well constrained, and that the temperature asymptotes to $11.1\pm
1.6$keV at large radii. Previous measurements of the temperature of
A2029 at large radii \citep{mol99,whi00,irw01}, indicate a gas
temperature of $T_g \sim 5-8$~keV, suggesting a turnover in $T_g$
between $300-500$~\hkpc. Future work  incorporating data at larger
radius will thus require a more sophisticated temperature model.

We note that as we found in Paper 1, the ICM of A2029 is apparently a
single-phase gas at all radii. Even within a radius of $3\arcsec$, a single
{\sc apec} model provides an excellent fit to the data, and there is simply no
spectral evidence to support an additional component such as a second
temperature, a cooling flow model, or a power-law. The B\&M model provides a
very good approximation to the $T_g$ data except for underestimation of the
fourth data point. We cannot determine if this is a statistical fluctuation, a
systematic error in our analysis, or real structure in the $T_g$ profile. The
last case would introduce a correction factor to the mass in that region,
otherwise the temperature, and thus the mass profile appears quite smooth (see
\S \ref{subsec_sys}).

\subsection{Mass Data and Fitted Profile \label{subsec_mass}}

At each radius $i$, we have calculated the total enclosed
gravitating mass $M(<\bar{r}_i)$ according to eq. \ref{eq_mtot},
using various density and temperature model fits. The logarithmic
derivatives of $\rho_g$ and $T_g$ are evaluated at $\bar{r}_i$, and
$T_g(\bar{r}_i)$ is interpolated from the fitted $T_g$ model. We
estimate statistical errors $\sigma_{M_i}$ on the mass data as
follows: For each Monte Carlo simulation $j$ of the $\rho_g$ and
$T_g$ data (\S \ref{subsec_bin}), we obtain a pair of fitted
profiles ($\rho_{fit_j}$ and $T_{fit_j}$) from which we calculate a
set of mass values $M(<\bar{r}_i)_j$. We thus obtain 100 mass
values at each radius, from which we calculate the standard
deviation and report it as the ``1$\sigma$'' error.

For the reference pair of gas models (the cusp model for $\rho_g$
and the B\&M model for $T_g$), we present the total enclosed mass
of A2029 in the left panel of Figure \ref{fig_mass_compare}
(\emph{open circles}). The profile has a nearly constant slope,
with a slight flattening in the inner 3 data points. At our final
data point, we obtain a total enclosed mass of $9.15\pm0.25 \times
10^{13} h_{70}^{-1}~\msun$ within $260$\hkpc. The mass calculated
using the power-law $T_g$ model is also shown (\emph{open
squares}). We note that there is significant uncertainty in the
mass at the innermost data point: while the temperature error is
relatively high ($\sim10\%$), the slopes of both $\rho_g$ and $T_g$
are even less well-constrained. In fact, using the double-$\beta$
model $\rho_g$ fit obtains a flat slope at this point, resulting in
a negative mass value (see \S \ref{subsec_sys}).

To analyze the shape of the mass profile, we fit parameterized
models to the best fit mass values. To estimate errors on the
parameters of these mass models, we obtain a mass fit $M_{fit_j}$
for each simulated mass data set $M(<\bar{r})_j$. As above, we
therefore obtain 100 values of each parameter in a given mass
model, from which we calculate a standard deviation. We have
overlaid in Figure \ref{fig_mass_compare} (\emph{Left Panel})
power-law fits to both mass profiles. The reference mass data are
well-fitted by a power-law $M(<\bar{r}) \propto r^{\alpha_m}$, with
slope $\alpha_m=1.81\pm0.04$ (\emph{solid line}), over the entire
mass range. The mass data obtained from using the power-law $T_g$
model are slightly higher below $r=17$\hkpc, and are also
well-fitted by a power-law with $\alpha_m=1.76\pm0.03$
(\emph{dashed line}).

Although the power-law is a good visual fit to the mass data, we
have examined whether $\chi^2$ is reduced by instead using a broken
power-law (BPL) model:
\begin{equation}
M(<r) = \begin{cases}
A (r/r_b)^{\alpha_{in}}, &  r < r_b \\
A (r/r_b)^{\alpha_{out}}, & r \geq r_b
        \end{cases}
\label{eq_bpow}
\end{equation}
where $r_b$ is the radius of the break between inner and outer
slopes. If all the data points are included in the fit, then we
obtain $r_b = 5-14$~\hkpc, with $\alpha_{in} = 1.1-1.5$, and
$\alpha_{out}$ consistent with $\alpha_m$ given above (these
results assumed a cusp $\rho_g$ model, and either the B\&M or
power-law $T_g$ model). However, if the inner data point is
excluded, then the BPL fits are consistent with the single
power-law fits given above. Since the inner data point is subject
to large systematic error due to the chosen $\rho_g$
parameterization (as above, see also \S \ref{subsec_sys}), we can
conclude there is no evidence for a significant break in the
logarithmic radial mass profile.

To account for more gradual deviations from a power law, we have
also fitted the mass data with the \citet*[NFW hereafter]{nav97}
profile, $\rho(r)\propto [(r/r_s)(1+(r/r_s)^2)]^{-1}$, the
\citet{her90} model, $\rho(r)\propto [r(1+r)^{3})]^{-1}$, and the
\citet[M99 hereafter]{moo99} model, $\rho(r)\propto
[(r/r_s)^{1.5}(1+(r/r_s)^{1.5})]^{-1}$. Integrating these density
profiles obtains mass models for fitting; analytic forms can be
found in the literature \citep[see, e.g.,][]{kly01,loe02}. These
fits are presented in the next section.

\subsection{The Dark Matter Distribution\label{subsec_dm}}

Assuming the cluster to be dominated by DM (a point we argue
below), the fitted total mass profile corresponds to an implicit DM
density distribution. For any power-law fit, $\rho_{DM} \propto
r^{\alpha_{DM}}$, where $\alpha_{DM} = \alpha_m -3$. For the
power-law fit to the reference mass profile we therefore observe a
dark matter density slope of $\alpha_{DM}=-1.19\pm0.04$.

In Figure \ref{fig_mass_compare}, (\emph{right panel}) we again
show the total enclosed mass profile (\emph{open circles}). To
compare with other clusters, as well as theoretical expectations,
it is convenient to scale the radius in terms of the virial radius,
$\rvir$\footnote{The virial radius is taken to be the radius at
which the matter density is 200 times the critical density required
for closure of the Universe.}. We calculate one popular predicted
value of $\rvir$ as a function of emission-weighted global
temperature, using the form given by \citet{neu99}, eq. 9, normalized to
the 2.0-9.5 keV band Mass-Temperature relation given by
\citet{mate01}, which we convert to our cosmology. If we
extract one spectrum in this band from the entire region within $186\arcsec$
which we observe with \chandra, we measure a single-temperature fit
of $7.54\pm0.15$~keV\footnote{Our emission-weighted temperature measurement
is in good agreement with the \beppo{} analyses of both \citet{irw00} 
and \citet{deg02}}. We thus obtain 
$\rvir = 2.71\pm0.42$\hmpc{} 
for A2029.
We have
plotted the upper axis of Fig. \ref{fig_mass_compare} (\emph{right
panel}) in units of $\rvir$, which shows that we are examining the
dark matter profile on a scale from $< 0.001-0.1 \rvir$.

We have overlaid fits to the mass profile of A2029 from three
different mass models: a power-law (\emph{dashed line}), an NFW
mass model (\emph{solid curve}), and an M99 model (\emph{dotted
curve}). While the power-law model provides a good overall fit
($\chi^2$/dof $=24.1/5$), the NFW model is preferred
($\chi^2$/dof $=11.8/5$). The data are more closely approximated by
the NFW profile except for a $\sim 1.5\sigma$ difference at the
innermost data point, which has additional systematic uncertainties
(as noted above) rendering the discrepancy insignificant (see \S
\ref{subsec_sys}). The Hernquist profile fit (not shown) is nearly
identical to the NFW fit ($\chi^2$/dof $=11.4/5$). The M99 model
does not provide an acceptable fit to the overall profile
($\chi^2$/dof $=250.1/5$), though its small radius slope is
compatible with the inner 3 data points. However, if fit solely to
these points, it falls well below the remainder of the mass
profile, which is better constrained. We note that although the NFW
and Hernquist models improve $\chi^2$ significantly, the relative
differences between these models vs. the power-law mass model are
small ($<10\%$ between 17 and 260~\hkpc). Unlike the BPL model,
which presents a sharp break, the gradual change in the logarithmic
slope of the NFW or Hernquist profiles better quantifies the small
deviations of the mass data from a pure power-law.

For the NFW profile, we find a scale radius $r_s = 540\pm 90
$\hkpc{}, and a concentration parameter 
$c=4.4\pm 0.9$. 
This allows
us to calculate the value of $\rvir$ expected from the NFW model
($\rvir \equiv c r_s$), for which we obtain
$\rvir$(NFW)$=2.39\pm0.62$\hmpc, in good agreement with $\rvir$ as
predicted by the Mass-Temperature relation given above.

Several other authors find that CDM halos (e.g., NFW or M99
profiles) are also consistent with their X-ray cluster observations
at $r\gtrsim 0.1\rvir$
\citep[e.g.,][]{ett02a,pra02,sch01,all01,ara02}, while
gravitational lensing studies report conflicting results in cluster
cores \citep[i.e.,][]{san02,nat02}. The Hydra A cluster exhibits
$\rho_{DM}\propto r^{-1.3}$ over a similar range of radii to our
analysis \citep{dav01}, however, there are prominent interactions
between the X-ray gas and the radio source in that system, implying
significant deviations from hydrostatic equilibrium. Thus the DM
profile inferred for A2029 (which is comparatively undisturbed in
this regime), provides evidence that the Hydra A results are
robust, and is an important independent confirmation of CDM
predictions.

\subsection{Possible Systematic Errors in the Mass
Profile\label{subsec_sys}}

We have explored several alternative data reduction and analysis
choices to investigate systematic errors in the mass profile. We
have calculated density, temperature and mass profiles in four
different sets of annuli: (1) the original 11 annuli used in Paper
1, (2) 13 annuli, obtained by dividing the inner 2 annuli of Paper
1 into 4 smaller annuli, (3) 5 annuli, using much larger bins, (4)
7 annuli, using a combination of the very small annuli from (2) in
the center with the larger, higher S/N annuli from (3). Binning
choice (4) is the reference choice for this paper, which shows the
most detail in the core, and provides the highest S/N for the
temperature estimates at larger radii. For all 4 choices of
binning, the resulting mass profiles are not perceptibly different
from the reference results. We have also analyzed the data in
energy ranges of $0.5-8.0$, $0.7-8.0$, and $1.0-8.0$ keV, and
repeated the entire experiment using a MEKAL plasma model rather
than the APEC reference model; in all cases the slope of the fitted
mass profile is within the $1\sigma$ errors of the reference fit.

The effect on the mass profile of our choice of eq. \ref{eq_re} to
calculate $\bar{r}$ can be tested by instead using 2 extreme cases
for the estimate of $\bar{r}$: assumed density slopes of $\rho_g
\propto r^0$ and $\propto r^{-3}$, which easily bracket all the
instantaneous slopes of the observed density profile \citep[see
also][]{ett02a}. We find that the mass profile does not vary
perceptibly with the choice of $\bar{r}$.

Our reference analysis is to use the same Period C background files
as in Paper 1. Variations in the X-ray flux of the extragalactic
sky between our target and the background templates may render them
inaccurate. We experimented with modifying the background flux by
changing the effective exposure time of the templates to $\pm20\%$
of its nominal value. While this can affect the estimated
temperature in the outermost annulus, the overall temperature
profile does not vary beyond the $1\sigma$ errors, nor does the
slope of the mass profile. We have corrected the data using the
``corrarf'' routine (\S \ref{sec_obs}), which results in lower
$T_g$ values than we reported in Paper 1, by up to 16\% in the
outermost bin. This has the effect of slightly flattening the
fitted $T_g$ profile, as well as the mass profile (the uncorrected
profile has overall slope $\alpha_m = 1.91\pm0.03$). The
calibration of the ACIS instrument below 1~keV is ongoing, but it
is reportedly now accurate at the $\approx 10\%$ level (see
footnote \ref{foot_cor}).

In Figure \ref{fig_mass_compare} (\emph{Left Panel}) we show the
total mass values obtained from each of the two temperature
profiles shown in Figure \ref{fig_rho_t} (in both cases, we use the
cusp fit to $\rho_g$). The mass values all overlap in their
$1\sigma$ error bars, as do the slopes of the mass profiles. We
have also explored the choice of the model which is fit to
$\rho_g$. The overall power-law fits to the mass profiles derived
from the cusp and double-$\beta$ profiles are indistinguishable.
However, we note that the double-$\beta$ model results in a
negative (and unphysical) mass value in the central bin (due to its
large statistical uncertainty, it does not affect the overall
profile fit).  The single $\beta$ model (which is an unacceptable
fit to the $\rho_g$ data) also obtains a very flat slope for the
inner 3 data points which results in negative mass values over that
region. To obtain a positive mass, eq. \ref{eq_mtot} requires a
negative sum for the logarithmic derivatives: inspection of Figure
\ref{fig_rho_t} shows that the slopes of neither $\rho_g$ nor $T_g$
are well-constrained at the radius of the innermost data point. The
cusp and double-$\beta$ models diverge in this regime, but the data
do not distinguish between them. If the gas was in fact isothermal
at this radius our mass estimate would be positive for all 3
$\rho_g$ fits, but without much higher S/N data (at higher spatial
resolution) we cannot infer such a state for the gas. \emph{We note
that exclusion of the innermost data point has no perceptible
affect on the mass profile fit.}

The use of parameterized functions for $\rho_g$ and $T_g$ has the
effect of smoothing those data, in turn resulting in a smoother
mass profile (\S \ref{subsec_profiles}). We have also calculated
the mass profile by using a simple ``point-to-point'' estimation of
the $\rho_g$ and $T_g$ slopes where the logarithmic derivatives
needed for eq. \ref{eq_mtot} are taken to be the slope between each
pair of adjacent points. The mass values are significantly
different only in the fourth and fifth bins, which lie above and
below the reference data points, respectively. Nonetheless, the
average slope of the mass profile is not altered beyond the
$1\sigma$ limits of the reference fit (as one would expect, since
the reference $\rho_g$ and $T_g$ models provide good fits to the
data). Aside from the innermost data point, the slopes of $\rho_g$
and $T_g$ are well constrained regardless of the parameterization,
and the overall mass profile is robust.

We have assumed spherical symmetry for A2029, though its mass
distribution exhibits an average ellipticity of 0.5 \citep{buo96}
within $\approx 1.6$\hmpc{} inferred from the X-rays and optical light.
However,  we have measured the spherically averaged mass
distribution, which is quite insensitive to the ellipticity of the
system (including the specific case of A2029, \citealt{buo96}; see
also \citealt{evr96}). This approach is also useful for comparison
with simulations which also present a spherically averaged profile
(i.e., NFW, M99).

\section{Gas Mass, Baryon Fraction, and $\Omega_{m}$
\label{sec_gas}}

We have calculated the total gas mass $M_{gas}(\bar{r})$ in each
spherical shell directly and summed them to obtain an enclosed gas
mass profile (i.e., we do not interpolate the $\rho_g$ values from
the fitted profile as we did for $T_g$ in the total mass
calculation; the results with either method are unchanged). In
Figure \ref{fig_mass_compare} (\emph{Right Panel}) we have overlaid
the total enclosed gas mass profile (\emph{open diamonds}). We see
that the gas mass at the center of the cluster is $<3\%$ of the
total mass, rising to $13.9\pm 0.4\%$ at our last measured data
point. This may safely be regarded as a lower limit, as simulations
and other measurements suggest a higher asymptotic value for the
gas mass fraction in clusters, f$_{gas}(\rvir) \sim
0.2-0.3h_{70}^{-3/2}$ \citep[e.g.,][]{all02}, while we only have
observations at $r\leq 0.1\rvir$.

Assuming that A2029 contains proportions of dark matter and
baryonic matter equal to their Universal values, one may place an
upper limit on $\Omega_{m}$ with f$_{gas}$. Given the Universal
baryon mass density from big bang nucleosynthesis calculations and
recent observations of the deuterium abundance ($\Omega_B$(BBN)),
one finds $\Omega_{m} = \Omega_B\rm{(BBN)}/{\rm{f}}_B$, where f$_B$
is the Universal baryon mass fraction. Given $\Omega_Bh^2\rm{(BBN)}
= 0.020\pm0.001$ \citep[see, e.g.,][]{bur01}, $h=0.7\pm0.07$, and
f$_B \geq$ f$_{gas} \geq 0.139 \pm 0.004h_{70}^{-3/2}$ from this
work, we estimate $\Omega_{m}\leq 0.29\pm 0.03h_{70}^{-1/2}$, in
agreement with current estimates \citep[for a recent review,
see][]{tur02}.

\section{Mass Budget of the Core\label{sec_dis}}

Massive elliptical galaxies such as NGC 720 \citep{buo02}, and NGC
4636 \citep{loe02}, are apparently DM-dominated all the way into
their cores. This situation is likely to exist in a cD cluster such
as A2029, except at possibly the smallest scales ($r\approx
1-10$kpc), where the cD stellar mass component may be more
important \citep{dub98}. We have found that the gas in A2029 is
only a few percent of the total mass in the core of the system. But
are we observing the stellar mass (e.g., the cD)?
\citeauthor{dub98} suggests the turnover between stellar and
DM-domination should occur at a few tens of kpc; but although we
are specifically observing this regime, we see no significant break
in the mass profile (see \S \ref{subsec_mass}). The outlying fourth
$T_g$ point and the central $\rho_g$ peak over a single $\beta$-model 
are intriguing, but the
balance of the data suggest that any transition from
stellar-dominated to DM-dominated regimes is quite smooth in this
system.

\subsection{Stellar Mass\label{subsec_stellar}}

We have estimated the contribution of stellar matter to the system,
by using optical observations of the cD galaxy in A2029 and its
diffuse halo. \citet{uso91} conducted a detailed optical analysis
of A2029, and observed that the surface brightness of the cD+halo
was well-fitted by a \citet{dev48} profile out to
$\approx300\arcsec$. They report a total $R$-band luminosity of
$5\times10^{11} h^{-2}_{100} \lsun$ for the cD+halo, which is 41\%
of the total light of the cluster within $260\arcsec$ radius. We
assume the cD+halo to be the vast majority of the light within its
effective radius $R_e = 52\arcsec$ (76\hkpc).  Using the
\citet{her90} model as an approximate deprojection of the $R^{1/4}$
law, we normalize to the total light of \citet{uso91}, converting
to our cosmology, and the $V$ band, and obtain a luminosity
profile. In Figure \ref{fig_ml}, we present the total cluster
mass-to-light ratio ($M_{tot}/L_V$) profile of the core of A2029,
based on the cD+halo light profile. It is consistent with a
constant $M_{tot}/L_V \approx12\msun/\lsun$ within 20\hkpc, and
rises rapidly at larger radius.

\begin{\myfigure}
\centerline{\epsfig{file=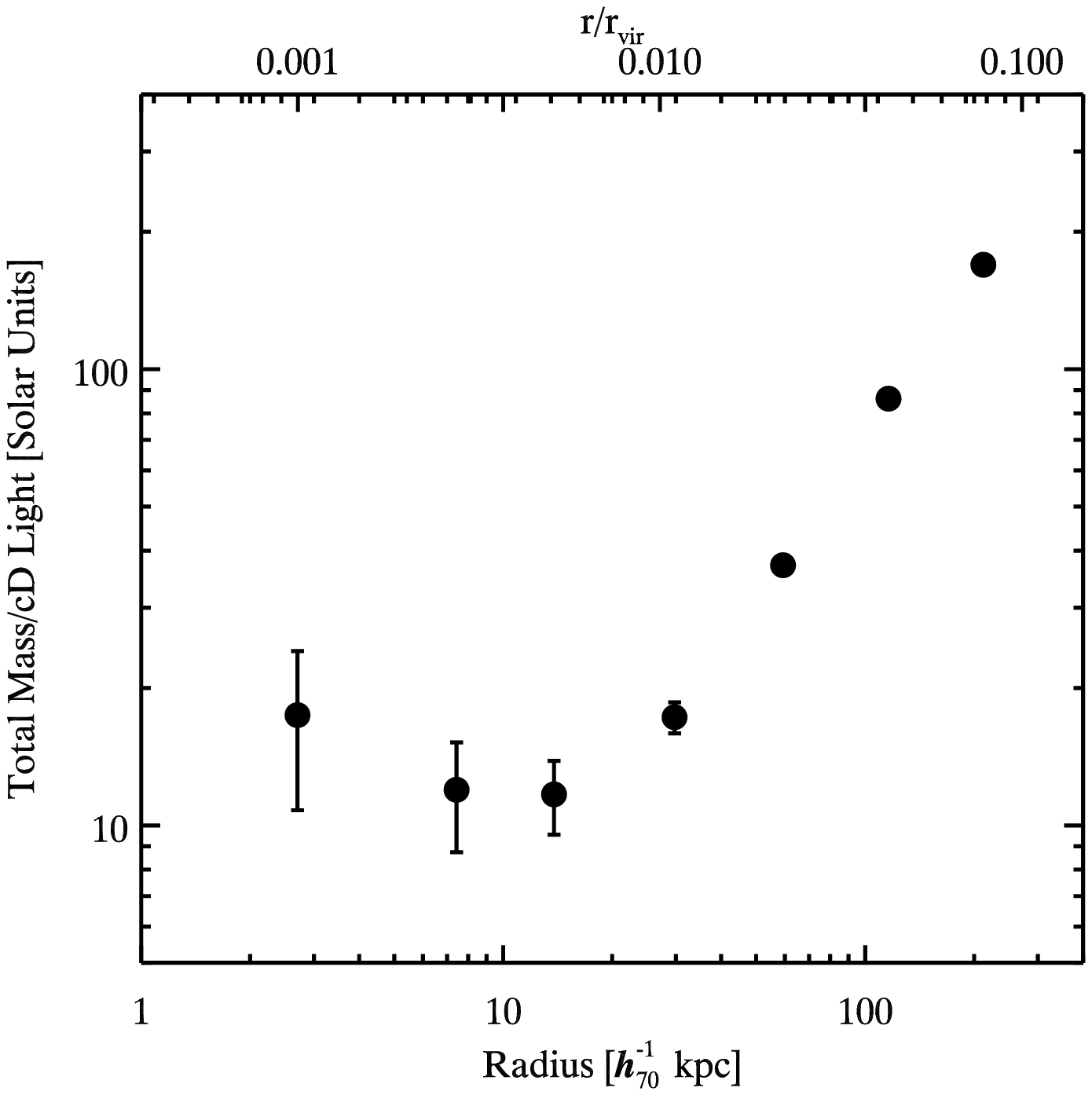, scale=0.65}}
\caption{Ratio of total enclosed cluster mass to stellar light from
the cD and diffuse envelope in A2029. The light follows an $R^{1/4}$
law with $R_e = 52\arcsec$, which we have approximated by a 
\citet{her90} profile normalized to a total cD+envelope $R$-band
luminosity of $5\times10^{11} h^{-2}_{100}\rm{L}_{\sun}$
\citep{uso91}, converted to our cosmology and the $V$-band. Error
bars represent the uncertainty in the mass estimate.
}
\label{fig_ml}
\end{\myfigure}
\smallskip

Dynamical stellar analyses estimate the total masses of galactic
systems, but cannot directly measure the stellar mass. Attempts to
resolve this with population synthesis techniques yield highly
uncertain results for the stellar mass-to-light ratio, $M_*/L_V =
1-12 \msun/\lsun$ \citep{pic85}. Using these bounding values for
$M_*/L_V$ we plot in Figure \ref{fig_mass_compare} (right panel)
associated stellar mass profiles (\emph{dot-dashed curves}) for the
cD+halo system. The data indicate that at $r<0.005\rvir$
($\approx15$\hkpc) the stellar mass can potentially dominate the
system, depending on the highly uncertain assumed $M_*/L_V$. The
stellar material outweighs the gas mass up to a radius between
$\approx 0.005-0.04\rvir$ ($\approx15-100$\hkpc). If we assert a
plausible value, $M_*/L_V = 5~\msun/\lsun$ for the system, we would
conclude that the cluster is DM dominated down to the smallest
scales measured here, which is consistent with a single NFW mass
component (and also with the lack of a significant break in the
total mass profile, \S \ref{subsec_mass}). Alternatively, the gas
density peak over the $\beta$-model at $r<17$\hkpc, 
or the outlying $T_g$ data point may
indicate the presence of a stellar mass component in excess of (or
differing from) an NFW DM halo.

\subsection{Velocity Dispersion\label{subsec_vel}}

An interesting comparison can be made with the optical velocity
dispersion profile of A2029. As noted by \citet{dub98}, IC1101 (the
cD galaxy in A2029) is one of the only brightest cluster galaxies
observed to have an increasing velocity dispersion, $\sigma_V$,
\citep{dre79}. We may regard the stars in the cD galaxy (from which
$\sigma_V$ is measured) to be simply another family of tracer
particles (as the hot ICM gas particles) bound to the same
gravitational potential.

Thus, the effective velocity of the ICM gas particles may be
expected to mirror that of the stars. We estimate the velocity
dispersion for the ICM gas as $\sigma_{ICM}^2 = kT_g/\mu m_p$.  The
temperature data yield $\sigma_{ICM}^2 \propto r^{0.27\pm0.01}$, while the
cD velocity dispersion profile of \citet{dre79}, measured over the
range $2-100 h_{50}^{-1}$~kpc reveals $\sigma_V^2 \propto
r^{0.25}$. This correspondence is not required for the condition of
hydrostatic equilibrium, but the similarity of the two species is
striking. A2029 is one of the few systems where the shape of both
the X-ray and optical velocity dispersions can be measured in
detail: the consistency between them is further evidence that we
are observing a dynamically relaxed system where all the mass
components are in equilibrium with the gravitational potential.

\section{Conclusions\label{sec_conc}}

We have analyzed high spatial resolution \chandra{} data of the
A2029 cluster of galaxies, obtaining well-constrained ICM gas
density and temperature profiles on scales of $0.001-0.1\rvir$
($\approx3-260$\hkpc). Fitting smooth functions to these profiles,
we obtain mass profiles for A2029, measuring the shape of the total
mass profile at unprecedented accuracy to a very small fraction of
the virial radius. Our results are insensitive to most details of
the data reduction, but are closely tied to the well-constrained
temperature and density distributions.

We find that the shape of the inferred dark matter density at
$<0.1\rvir$ is consistent with the NFW parameterization of CDM
halos, but apparently incompatible with that of M99, though we note
that individual objects may be expected to show significant scatter
from a mean DM halo profile \citep{bul01}.


The consistency of
the NFW model with the mass profile all the way
down to $<0.01\rvir$ indicates that there is no need to modify the
standard CDM paradigm to fit the DM distribution in this cluster,
consistent with X-ray observations of other clusters at larger radii
(see \S \ref{subsec_dm}). This result contrasts with the strong
deviations from the CDM predictions observed in the rotation curves
of low surface brightness galaxies \citep[e.g.,][]{swa00}, and dwarf
galaxies \citep[e.g.,][]{moo99} which inspired the self-interacting
DM model \citep{spe00} to explain the relatively flat core density
distributions in these galaxies.  Hence in light of the good
agreement with the NFW profile in clusters, the deviations observed
on small galaxy scales do not seem to imply a fundamental problem
with the general CDM paradigm. Instead, it is likely that the
numerical simulations do not currently account properly for the
effects of feedback processes on the formation and evolution of
small halos.

%
Our analysis suggests that A2029 is dominated by a single mass (i.e., DM)
component at all measured scales below $0.1\rvir$, or that any transition
from a stellar mass dominated component in the cD and a DM component
is quite gradual.
We also observe a rising gas fraction from $<3\%$ to $>14\%$ in
A2029, obtaining an upper limit to $\Omega_m \leq 0.29\pm0.03
h_{70}^{-1/2}$, consistent with other current studies.

\acknowledgments

The authors wish to thank Amit Lakhanpal for an initial calculation
of the mass profile. This work was supported by \chandra{} grant
G00-1021X. 
\bigskip
\\
\\



\end{document}